\documentclass[11pt,a4paper]{article}
\pdfoutput=1

\usepackage{jheppub}
\usepackage{rotating}

\usepackage{wrapfig}

\usepackage{amsfonts}
\usepackage{amsmath}
\usepackage{slashed}
\usepackage{amssymb}
\usepackage{latexsym}
\usepackage{multirow}
\usepackage{hyperref}
\usepackage{enumitem}
\hypersetup{colorlinks=true,citecolor=red,linkcolor=magenta,urlcolor=blue}
\usepackage[caption=false]{subfig}
\usepackage{relsize}
\usepackage{booktabs}
\usepackage{cleveref}
\usepackage{fancyhdr}
\usepackage{float}
\usepackage{graphicx}
\usepackage{microtype}
\usepackage{tabularx}
\usepackage{xcolor}
\usepackage[bottom]{footmisc}

\title{IRC-safe Graph Autoencoder for unsupervised anomaly detection}
\author[a]{Oliver Atkinson,}  
\author[a]{Akanksha Bhardwaj,}
\author[a]{Christoph Englert,} 
\author[b]{Partha Konar,}
\author[b,c]{Vishal S. Ngairangbam,}
\author[d]{Michael~Spannowsky}

\affiliation[a]{School of Physics \& Astronomy, University of Glasgow, Glasgow G12 8QQ, United Kingdom}
\affiliation[b]{Theoretical Physics Division, Physical Research Laboratory, Shree Pannalal Patel Marg, Ahmedabad, 380009, Gujarat, India}
\affiliation[c]{Discipline of Physics, Indian Institute of Technology, Palaj, Gandhinagar - 382424, Gujarat, India}
\affiliation[d]{Institute for Particle Physics Phenomenology, Durham University, Durham DH1 3LE, United Kingdom}

\emailAdd{o.atkinson.1@research.gla.ac.uk} 
\emailAdd{akanksha.bhardwaj@glasgow.ac.uk}
\emailAdd{christoph.englert@glasgow.ac.uk}
\emailAdd{konar@prl.res.in}
\emailAdd{vishalng@prl.res.in}
\emailAdd{michael.spannowsky@durham.ac.uk}
\abstract{Anomaly detection through employing machine learning techniques has emerged as a novel powerful tool in the search for new physics beyond the Standard Model. Historically similar to the development of jet observables, theoretical consistency has not always assumed a central role in the fast development of algorithms and neural network architectures. In this work, we construct an infrared and collinear safe autoencoder based on graph neural networks by employing energy-weighted message passing. We demonstrate that whilst this approach has theoretically favourable properties, it also exhibits formidable sensitivity to non-QCD structures.}

\begin{document}
\maketitle
\flushbottom
\section{Introduction}
\label{sec:intro}
New physics searches at the high-energy frontier of the Large Hadron Collider (LHC) have so far not resulted in any significant deviation of experimental results from the Standard Model (SM) expectation. However, with a growing dataset of these high energy measurements, the pressure mounts for theoretically motivated scenarios of beyond the SM (BSM) physics which have been devised to tackle known shortcomings of the SM. So what are the ways out of this juxtaposition of experimental agreement with the SM and its failure to describe established physics at small and large distances? 

On the one hand, there is an increasing emphasis on theoretically as-model-independent-as-possible approaches based on effective field theory (EFT)~\cite{Weinberg:1978kz}. EFT navigates QFT correlations away from the SM prediction in any possible direction given the SM symmetry and particle content, thus avoiding UV model biases. Alas, such an approach poses its own challenges: looking for deviations from the SM expectation along these lines involves many ad-hoc interactions. Concrete models will typically only source a subset of relevant interactions, e.g. \cite{Englert:2019xhk,DasBakshi:2020ejz,Bakshi:2021ofj}. There has been great progress to facilitate matching calculations~\cite{Carmona:2021xtq}, however, depending on the new physics scenario, this can create a significant overhead that must be included in the parameter fitting procedure itself~\cite{Freitas:2016iwx,Englert:2019rga}.

On the other hand, we can look for phenomenological deviations from specific SM signatures directly in collider results without any new physics bias. Under the assumption that collider data can be modelled sufficiently adequately, we can employ the SM expectation to identify regions where measurements do not follow the SM expectation. This anomaly detection has emerged as a powerful tool to look for any hidden signature of new physics in the data. Recently, a range of state-of-the-art methods for anomaly detection~\cite{Atkinson:2021nlt,Blance:2020ktp,Collins:2021nxn,Aaboud:2018ufy,Collins:2018epr,Blance:2019ibf,Hajer:2018kqm,DeSimone:2018efk,Araz:2021wqm,Nachman:2020lpy,Hallin:2021wme,Nachman:2020ccu,Cheng:2020dal,Canelli:2021aps} using deep learning have been designed. 

Theoretical consistency when confronting collider data with theoretical expectations is pivotal. The formulation of infrared and collinear (IRC)-safe observables is necessary to guarantee the comparability of experimental measurements and theoretical predictions to all orders in perturbation theory employing the Kinoshita-Lee-Nauenberg (KLN) theorem~\cite{Kinoshita:1962ur,Lee:1964is} and collinear factorisation~\cite{Collins:1989gx} of parton distributions. Any sensitivity enhancement observed by algorithms that inadvertently employ IRC-unsafe information will be critically assessed in subsequent studies, with a potentially meaningless theoretical outcome. This is a tedious task all too familiar from the use of IRC-unsafe jet clustering algorithms (e.g. iterative cone algorithms) by the CDF and D0 experiments~\cite{Kilgore:1996sq}, which was later only partially addressed with the midpoint algorithm during Tevatron Run II, until fully IRC-safe algorithms~\cite{Catani:1993hr,Cacciari:2008gp} were established as the only theoretically meaningful community consensus.

It seems prudent to avoid mistakes of the past: in this paper, we devise an IRC-safe Graph Neural Network (GNN) autoencoder algorithm, employing an Energy-Weighted Message-Passing Network (EMPN)~\cite{Konar:2021zdg} for unsupervised anomaly detection. While the IRC-safe loss function is the primary observable of our autoencoder, we also study the latent space (graph) representation structure as a motivating tool for new physics discrimination~\cite{Dillon:2021nxw,Atkinson:2021nlt} and highlight the relations to known and more ``traditional'' IRC-safe observables. This paper is organised as follows: in Sec.~\ref{sec:outl}, we outline our EMPN approach and detail our IRC-safe graph construction before we introduce the IRC-safe autoencoder architecture and simulation framework in Sec.~\ref{sec:ircgraph}. Sec.~\ref{sec:results} is devoted to the discussion of the sensitivity performance of the autoencoder; we also highlight the correlation of sensitivity with more traditional jet-based observables. We conclude in Sec.~\ref{sec:conc}.

\section{A brief outline of Energy-Weighted Message Passing algorithm}
\label{sec:outl}
This section presents a brief overview of the IRC safe Energy-weighted Message passing algorithm~\cite{Konar:2021zdg}. It generalises Energy Flow Networks~\cite{Komiske:2018cqr,Dolan:2020qkr}, an IRC safe feature extraction on point clouds, by learning relational information between two elements (nodes) by constructing a graph out of the point cloud. This procedure is similar to message-passing networks like the Dynamic Graph Convolutional Neural Network (DGCNN)~\cite{wang2019dynamic} that extract local features beyond the global feature extraction via point-cloud-based architectures such as deep-sets~\cite{zaheer2017deep} and {\tt PointNet}~\cite{charles2017pointnet,qi2017pointnet++}. The algorithm consists of two necessary ingredients:
\begin{itemize}
	\item an IRC safe prescription for constructing graphs which guarantees that the graph is invariant under soft and collinear splittings; 
	\item an energy-weighted summed aggregation of messages (and node features after the final message-passing layer) taking the directional (unit vectors or angles) inputs $\hat{p}_i$ and $\hat{p}_j$ of the nodes connected by an edge $(j,i)$ at the initial layer.
\end{itemize}
In the following, we discuss these two elements separately.

\subsection{IRC safe graph construction}

\begin{figure}
        \centering
	\includegraphics[width=0.96\textwidth]{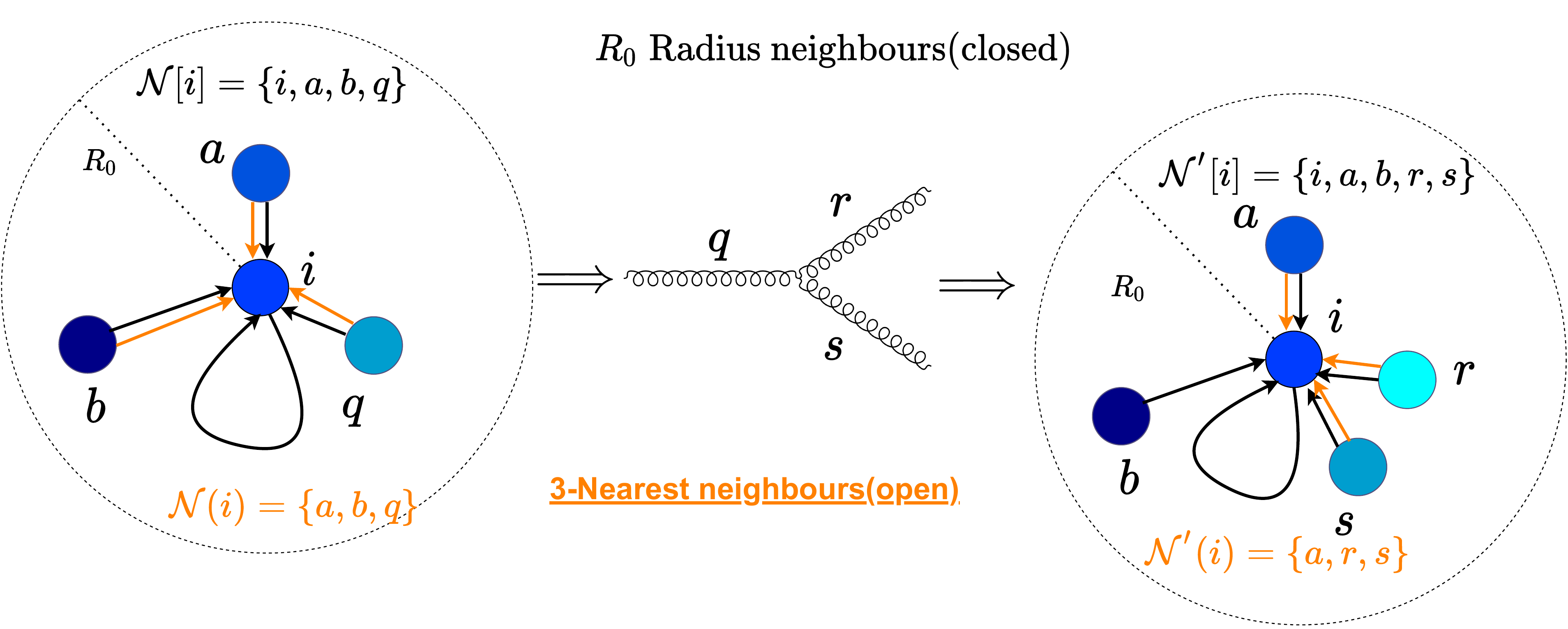}
	\caption{Representation of radius graph with $R_0$ in the $(\eta,\phi)$ plane undergoing a QCD splitting. The black arrows correspond to the connections of a radius graph, while the red arrows highlight the 3-nearest neighbours connections. One can see that the radius neighbourhoods have the same total energy, which is not the case for those obtained by the nearest neighbours method, leading to an IRC-unsafe construction.	\label{fig:R0_graph}}
\end{figure} 
The inductive biases that a message-passing algorithm imposes on its input data are highly dependent on the graph structure. For instance, the neighbourhood sets (the set of particles a node is connected with) determine the local connectivity of the nodes. Thus, the graph construction algorithm from a point cloud gives a strong indication that a graph neural network is the best avenue to pursue. The IRC safety of a message-passing algorithm also depends on the graph construction, and we highlight such an IRC safe graph construction algorithm in this section. 

Let $\mathcal{S}=\{p_1,p_2,p_3,.....,p_N\}$ be the set of four-vectors of the particles within a jet, while $\mathcal{S}'=\{p_1,p_2,....,p_{N+1}\}$ is the same set in the presence of an additional splitting. The collinear limit is when the emitted particles $r$ and $s$  with the angular separation $\Delta_{rs}$ tending to zero, while the soft limit refers to the case when one of the particle's energy tends to zero.
These four-vectors can be written as 
\begin{equation}
\label{eq:hat}
p_i=(z_i,\hat{p}_i)\,,~\hbox{with}~z_i={p^i_T\over \sum_{j\in\mathcal{S}} p^j_T }~\hbox{and}~\hat{p}=(\eta,\phi)\,,
\end{equation}
for hadron colliders, with the separation in the $\eta-\phi$ plane between two particles $i$ and $j$, defined as $\Delta R_{ij}=\sqrt{\Delta \eta_{ij}^2+\Delta\phi_{ij}^2}$ denoting the quantity analogous to $\Delta_{ij}$. Since we will be taking directed edges, the neighbourhood set of a node $i$ will be the set of all nodes with incoming connections to $i$. For all particles $i$ in $\mathcal{S}$ or $\mathcal{S}'$, a graph construction algorithm will construct neighbourhood sets $\mathcal{N}[i]$ 
and $\mathcal{N}'[i]$, respectively. We will use a ``closed'' neighbourhood with $i\in\mathcal{N}[i]$ instead of an ``open'' neighbourhood $i\notin\mathcal{N}(i)$, since the second choice will always be IRC unsafe when the node $i$ splits. To illustrate this, we show the radius graph with $R_0$ in the $(\eta,\phi)$ plane in Fig.~\ref{fig:R0_graph}, where the node $q$ undergoes a splitting. The black arrows highlight the connections of the radius graph. Fig.~\ref{fig:R0_graph} also demonstrates a nearest neighbourhood connection as an example of an IRC unsafe graph construction. 

To formalise the graph construction algorithm in terms of the four-vectors of the particles, we define a decision function $\mathbf{D}(p_i,p_j)$ and a threshold function $\mathbf{T}(p_i,p_j)$, such that any particle $j$ with four-vector $p_j$ will be assigned to the neighbourhood of particle $i$ with four-vector $p_i$ if $\mathbf{D}(p_i,p_j)$ is less-than or equal-to $\mathbf{T}(p_i,p_j)$. This can be summarised as
\begin{equation}
\label{eq:graph_def}
\mathbf{D}(p_i,p_j)\leq\mathbf{T}(p_i,p_j)\implies j\in\mathcal{N}[i]\,.
\end{equation}
Since we are interested in the soft and collinear limits, constructing an IRC safe graph requires putting conditions on these functions in the respective kinematical configurations.

The required condition on these functions for a ``parent'' splitting $q\to r+s$ when the ``daughters'' $r,s$ become collinear is 
\begin{equation} 
\label{eq:DT_c_split_dest}
\begin{split}
\mathbf{D}(p_i,p_r+p_s)\leq \mathbf{T}(p_i,p_r+p_s)\Leftrightarrow \mathbf{D}(p_i,p_r)\leq \mathbf{T}(p_i,p_r)\;\land&\;\mathbf{D}(p_i,p_s)\leq \mathbf{T}(p_i,p_s)\,,\\
\mathbf{D}(p_r+p_s,p_i)\leq \mathbf{T}(p_r+p_s,p_i)\Leftrightarrow \mathbf{D}(p_r,p_i)\leq \mathbf{T}(p_r,p_i)\;\land&\;\mathbf{D}(p_s,p_i)\leq \mathbf{T}(p_s,p_i)\,,
\end{split}  
\end{equation} where the second condition arises since the nodes $q,\;r$ or $s$ can also be the node whose neighbourhood is being determined. The only requirement in the IR limit for a daughter particle is that all the particles in the set $\mathcal{N}[i]$ are also present in $\mathcal{N}'[i]$, with the only potential addition of a soft particle. This is guaranteed by the form of Eq,~\eqref{eq:graph_def}, since both functions depend only on the four-vector of the two nodes of interest.\footnote{This is not the case for popular graph construction algorithms like $k$-nearest neighbours, for which the decision and threshold has a complicated dependence on the distance of the primary node $i$ with every other particle in the graph, and on the number of elements in the neighbourhood set.}
The conditions (c.f. Eq.~\ref{eq:DT_c_split_dest}) are satisfied in the collinear limit $\Delta_{rs}\to0$ if 
\begin{equation}
\mathbf{D}=\mathbf{D}(\hat{p}_i,\hat{p}_j)\,,\qquad \mathbf{T}=\mathbf{T}(\hat{p}_i,\hat{p}_j)\,,
\end{equation}
employing the definitions Eq.~\eqref{eq:hat}. Therefore, graphs formed by connecting particles within a constant radius $R_0$  in the $\eta-\phi$ plane are IRC safe when the decision and threshold functions take the form 
\begin{equation}
\label{eq:use_DT}
\mathbf{D}=\Delta R_{ij}\,,\qquad \mathbf{T}=R_0\,.
\end{equation} 
Note that these choices of $\mathbf{D}, \mathbf{T}$ yield closed neighbourhoods without additional requirements. We will use these graphs in the remainder of this paper; the neighbourhood of a particle of such a radius graph is shown in Fig.~\ref{fig:R0_graph}. 

\subsection{Energy-weighted message passing}
We detail the IRC safe message passing operation in this section. Before doing so, we summarise the general definition of message passing operation in the following steps. The first step, the message-passing stage, involves calculating the messages for all edges present in the graph. The message function, parametrised as a multilayer perceptron shared for all edges, takes the node features of the two nodes connected by an edge and evaluates the message. Since the message function does not need to be symmetric for the two node features, a direction convention is necessary for the second phase. In our convention, the message originates from all nodes in the neighbourhood $\mathcal{N}[i]$ and flows towards the particle $i$. 
	The second step, the node-readout stage, updates the node features of each node in the graph as a permutation-invariant function of all incoming messages.

IRC safety of the updated node features after a message-passing operation is crucially dependent on the nature of the node readout. A readout based on the maximum or minimum value of the node features depends on a single node feature in the neighbourhood, and a soft or collinear splitting of this particular node would render the updated node feature IRC-unsafe. 
This 
is ultimately 
related to 
identifying a specific node in the neighbourhood as special,\footnote{This is also the reason for using closed neighbourhoods $\mathcal{N}[i]$, as an open neighbourhood $\mathcal{N}(i)$, would give a special status to the node $i$.} which impedes KLN cancellations.
A mean readout, on the other hand, explicitly depends on the cardinality of the neighbourhood sets $\mathcal{N}[i]$ which is not a well-defined QCD quantity either since there can be an arbitrary but finite amount of resolvable emissions in the enhanced collinear or soft regions of phase space. Thus we use a summed readout, which will inclusively take all the particles in the neighbourhood into account and will not explicitly depend on their size.

An IRC safe graph construction algorithm ensures two things: the equality of the sum of energy (transverse energy in the case of hadron colliders) of all particles in either neighbourhood sets and the presence of both collinear daughters in $\mathcal{N}'[i]$ if the parent is present in $\mathcal{N}[i]$. Defining a scope-dependent energy weight-factors analogous to $z_i$ as
\begin{equation*}
\omega_j^{(\mathcal{K})}=\frac{p_T^j}{\sum_{k\in\mathcal{K}}\;p_T^k} \quad,
\end{equation*}with $\mathcal{K}$ denoting the set of particles in the particular readout operation, any message passing of the form
\begin{equation}
\label{eq:ewmp}
\mathbf{h}^{(l+1)}_i=\sum_{i\in\mathcal{N}[i]}\;\omega^{(\mathcal{N}[i])}_{j}\;\;\hat{\Phi}^{(l)}(\mathbf{h}^{(l)}_i,\mathbf{h}^{(l)}_j)\,,
\end{equation}
with $\mathbf{h}^{(0)}_i=\hat{p}_i$ and $\mathbf{h}^{(l)}_i$ denoting the updated node-features after $l$ message-passing operations satisfies IRC safety; 
in the infrared limit, it is straightforward to see that any soft particle with $z_r\to0\implies\omega_r^{(\mathcal{N}[i])}\to0$ for any node $i$. The splitting $q\to r+ s$ for IRC-safe graphs therefore yields
\begin{equation} 
\label{eq:w_sum}
\omega_q^{(\mathcal{N}[i])}=\omega_r^{(\mathcal{N}[i])}+\omega_s^{(\mathcal{N}[i])}\,.
\end{equation} 
In the collinear limit with $\hat{p}_q=\hat{p}_r=\hat{p}_s$ we have $\hat{\Phi}^{(0)}(\hat{p}_i,\hat{p}_q)=\hat{\Phi}^{(0)}(\hat{p}_i,\hat{p}_r)=\hat{\Phi}^{(0)}(\hat{p}_i,\hat{p}_s)$. 
Combining this with Eq.~\eqref{eq:w_sum}, we obtain (for $l=0$)
\begin{equation*} 
\omega_q^{(\mathcal{N}[i])}\;\hat{\Phi}^{(0)}(\hat{p}_i,\hat{p}_q)=\omega_r^{(\mathcal{N}[i])}\;\hat{\Phi}^{(0)}(\hat{p}_i,\hat{p}_r)+\omega_s^{(\mathcal{N}[i])}\;\hat{\Phi}^{(0)}(\hat{p}_i,\hat{p}_s)\,.
\end{equation*} 
When evaluating Eq.~\eqref{eq:ewmp} for the neighbourhood of a node $i$, the terms on the RHS and LHS of this expression are the only ones which will not be common between $\mathcal{N}[i]$ and $\mathcal{N}'[i]$, due to the IRC safe graph construction. The same expression is followed when $i=q$ on the left, and $i=r$ or $i=s$ on the right, since $\{r,s\}\subset\mathcal{N}'[s]$ and $\{r,s\}\subset\mathcal{N}'[r]$, with all three neighbourhoods (including $\mathcal{N}[q]$) containing the same particles except for $q$, $r$, and $s$. 
Therefore, from Eq.~\eqref{eq:ewmp} we have  $\mathbf{h}^{(1)}_q=\mathbf{h}^{(1)}_r=\mathbf{h}^{(1)}_s$ for collinear splittings. On the other hand, for a soft daughter, say $r$, we have $\mathbf{h}^{(1)}_q=\mathbf{h}^{(1)}_s$, but $\mathbf{h}^{(1)}_r \neq \mathbf{h}^{(1)}_q$, with $\mathbf{h}^{(1)}_r$ not necessarily zero. The presence of the node features of the daughter particles, even in the soft or collinear limit, impedes an IRC safe examination of the full jet graph unless observables are specifically designed to be insensitive to their presence in the IRC limit. The procedures  to take care of these additional nodes are explained in the following sections, which are different for supervised and unsupervised methods. Since the above derivation used the collinearity of $q$, $r$, and $s$, for IRC safe neighbourhoods, for the same neighbourhoods and any successive application of an energy-weighted message passing of the form Eq.~\eqref{eq:ewmp}, we have $\mathbf{h}^{(l)}_q=\mathbf{h}^{(l)}_r=\mathbf{h}^{(l)}_s$ for any $l$.

\section{IRC-safe graph autoencoder}
\label{sec:ircgraph}

In a supervised machine learning scenario, the IRC-safe graph readout acting on the node features of the final message-passing operation gives an IRC-safe graph representation, and one loses the graph's structure. The graph representation, a fixed-length vector obtained after applying a permutation invariant function on the node features for any variable-length graph, feeds into the downstream network. Therefore, training a classifier on the loss function defined with the downstream network's output proceeds without any complications from the presence of additional soft or collinear nodes. On the other hand, a graph autoencoder similar to the one proposed in Ref.~\cite{Atkinson:2021jnj} preserves the graph structure until the output. Therefore, the autoencoder's output graph will have additional nodes in the soft and collinear limits in the case of extra emissions. Since the observable of interest for anomaly detection with an autoencoder is the loss function, we need to ensure its IRC safety. In this section, we first devise an IRC safe loss function and give details of the network architecture and training.

\subsection{An IRC-safe loss function}
The definition of the loss function involves input which changes with a soft or collinear splitting. Therefore, the loss which is normally used as an observable  in anomaly detection, needs to be IRC-safe. A simple IRC-safe loss function for a jet with constituent set $\mathcal{G}$ is of the form
\begin{equation}
\label{eq:lNodeoss} 
\mathcal{L}_{\mathcal{G}}= \sum_{i \in \mathcal{G}} z_i ~d (\hat{p}_i,\hat{\bar{p}}_i)\,.
\end{equation}
The barred quantities are the output of the network, while the unbarred quantities are the inputs to the network.
The function $d (\hat{p}_i,\hat{\bar{p}}_i)\geq d_0$ denotes a well-behaved metric (one-to-one) between the input and the output space, with $d (\hat{p}_i,\hat{p}_i)=d_0$. We now show that this is indeed an IRC safe choice:

Any soft particle $s$, will not contribute to the sum since $z_s\to 0$, and hence it is IR safe. For the splitting $q\to r+s$ we have 
\begin{equation*}
\begin{split}
\mathcal{L}_\mathcal{S}&=...+z_q\;d(\hat{p}_q,\hat{\bar{p}}_q)+...\\
\mathcal{L}_\mathcal{S'}&=...+
z_r \;d(\hat{p}_r,\hat{\bar{p}}_r)+z_s \;d(\hat{p}_s,\hat{\bar{p}}_s)+...\quad.
\end{split} 
\end{equation*} 
Since, by construction, a GNN's node output after $L$ total message-passing operations $\mathbf{h}^{(L)}_i=\hat{\bar{p}}_i$, is a function of the input four-vectors $\{p_1,p_2,p_3,....p_N\}$, in general, they can have a very complicated dependence on all the input node features. However, due to the IRC safety of the EMPN, we have \begin{equation}
\label{eq:loss_diff}
\mathcal{L}_\mathcal{S'}-\mathcal{L}_\mathcal{S}=z_r \;d(\hat{p}_r,\hat{\bar{p}}_r)+z_s\;d(\hat{p}_s,\hat{\bar{p}}_s)-z_q\;d(\hat{p}_q,\hat{\bar{p}}_q)\,.
\end{equation}
In the collinear limit with $\hat{p}_q=\hat{p}_r=\hat{p}_s\implies \hat{\bar{p}}_q=\hat{\bar{p}}_r=\hat{\bar{p}}_s$, we therefore have (since $z_q=z_r+z_s$), 
\begin{equation}
z_q\;d(\hat{p}_q,\hat{\bar{p}}_q)=z_r \;d(\hat{p}_r,\hat{\bar{p}}_r)+z_s\;d(\hat{p}_s,\hat{\bar{p}}_s)\implies\mathcal{L}_\mathcal{S'}-\mathcal{L}_\mathcal{S}=0\,,
\end{equation}
i.e. collinear safety. In the following analysis of the EMPN autoencoder we will use mean-squared error between the input and output node features for $d (\hat{p}_i,\hat{\bar{p}}_i)$.

\subsection{Jet graph definition} 
To demonstrate the performance of the described algorithm, we use the publicly available top-tagging dataset of Refs.~\cite{kasieczka_gregor_2019_2603256,Butter:2017cot}. The dataset contains a training, validation and testing set of 600k, 200k, and 200k QCD jets, respectively. The training and validation are done only with the background QCD samples since the total cross-section of their production would be orders of magnitude higher than most probable signals. Although the dataset has the same number of top jets for each of the three analysis stages, we use the 200k top jets of the test dataset as a benchmark signal scenario. These jets are simulated using {\tt{Pythia8}}~\cite{Sjostrand:2014zea,Sjostrand:2007gs} and passed through {\tt{Delphes3}}~\cite{deFavereau:2013fsa} for the detector simulations using the default ATLAS parameter card. Jets are clustered from particle flow (Eflow) constituents with a distance parameter $\Delta R =0.8 $ using the anti-$k_t$ algorithm~\cite{Cacciari:2008gp}. The transverse momentum of the jets is in the range $p_T\in[550,650]$~GeV.  

Using the constituents of these jets, we construct the radius graphs which serve as the input to the IRC safe graph network. To construct the jet radius graph, we first calculate the inter-particle distance $\Delta R_{ij}$ in the $(\eta,\phi)$ plane. Next, we define a set of all the particles $i$ as the neighbourhood set $\mathcal{N}[i]$ such that $\Delta R_{ij}\leq R_0$, where $R_0$ is an external tunable parameter. Each node is associated with three node features
\begin{equation}
\label{eq:node_feat}
\mathbf{h}^0_i=({\Delta \eta_i,\Delta \phi_i,\Delta R_i})\,,
\end{equation}
where $\Delta \eta_i$, $\Delta \phi_i$, $\Delta R_i$ are calculated with respect to the jet axis.  For the network analysis, we choose $R_0 = 0.3$. Since the dependence of the classification power on $R_0$ for the supervised case was found to be mild~\cite{Konar:2021zdg}, with the AUC values changing in the third decimal value for different values of $R_0$ between $0.1$ and $0.5$, we restrict ourselves to a single value in the intermediate range.  The final node vectors  contain information about the $L$-hop neighbourhood with an effective radius of $R_0\times L$. On the other hand, the primary region of activity for the one-prong QCD jets used to train the network lies in a relatively small central region of the total jet of radius $\Delta R=0.8$. Therefore, the features learnt by the autoencoder would be weakly dependent on $R_0$, once the effective radius covers a significant portion of the central region.

\subsection{Network architecture and training}

\begin{figure}[!t]
	\centering
	\includegraphics[width=\textwidth]{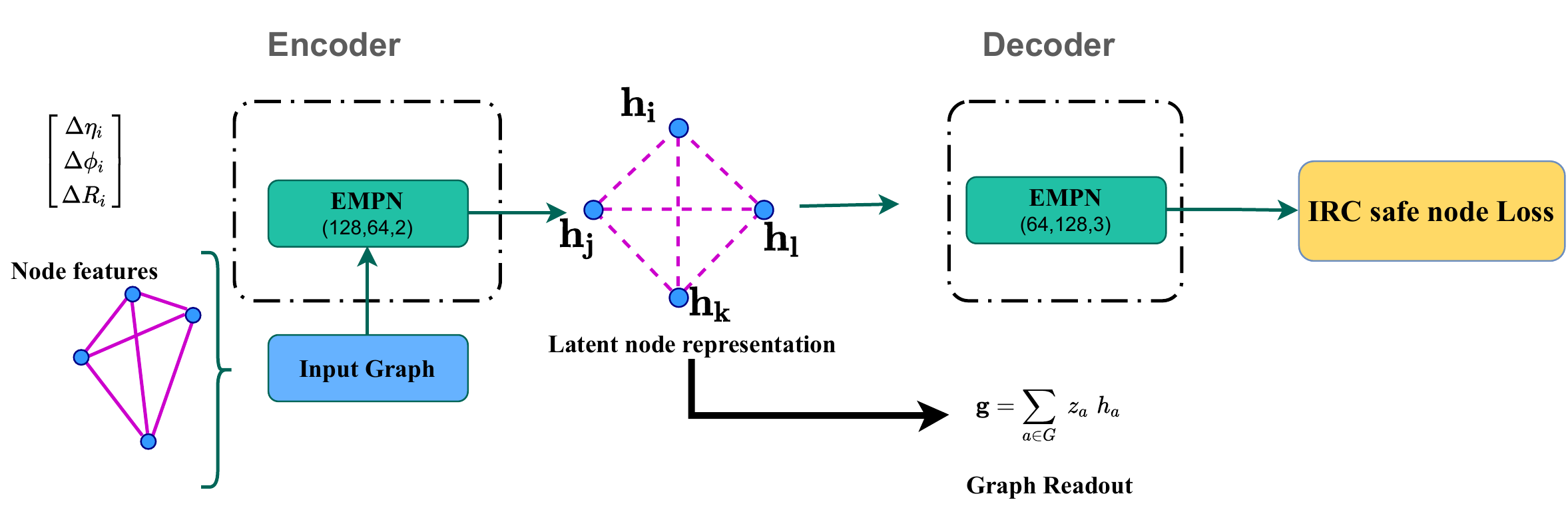}
	\caption{A schematic diagram of an IRC safe graph-autoencoder.	\label{fig:Network_arch_IRC}}
\end{figure}

Now that we have described the construction of the jet graphs, we discuss the details of the network architecture and training in this section. Follow from Fig.~\ref{fig:Network_arch_IRC} where we sketch a schematic diagram of an IRC safe graph-autoencoder. The encoder consists of three edge convolution operations with output dimensions of 128, 64 and 2, which is the dimension of the latent representation. Since we take three-dimensional node features, we restrict ourselves to a 2-dimensional latent space $(g_1,g_2)$ to induce an information bottleneck.\footnote{The effective number of inputs to a message function could be twice the number of input features--one each for the two nodes connected by an edge. However, a concrete understanding of the universal approximation properties of graph neural networks~\cite{NEURIPS2020_e4acb4c8} is yet to be achieved, making it difficult to precisely determine the actual input dimensions when looking at the complete graph neural network.}
The decoder also has three edge convolution operations, with the first two dimensions mirroring the encoder network dimensions (excluding the latent dimension). Finally, the last edge convolution operation maps the 128-dimensional node vectors at the penultimate message passing the layer to a three-dimensional space to reconstruct the input node features. 

We take $\hat{\Phi}^{(l)}$ at each message-passing layer to be a multilayer perceptron (MLP). For an edge convolution operation, we have for two node features $\mathbf{h}^{(l)}_i$ and $\mathbf{h}^{(l)}_i$ connected by an edge in Eq.~\eqref{eq:ewmp}, 
\begin{equation*}
\hat{\Phi}^{(l)}(\mathbf{h}^{(l)}_i,\mathbf{h}^{(l)}_j)=\hat{\Phi}^{(l)}\left(\mathbf{h}^{(l)}_i\oplus(\mathbf{h}^{(l)}_j-\mathbf{h}^{(l)}_i)\right)\,.
\end{equation*} 
Therefore the input vector to the MLP has twice the node-feature's dimensions, since the direct sum $\mathbf{h}^{(l)}_i\oplus(\mathbf{h}^{(l)}_j-\mathbf{h}^{(l)}_i)$, is a concatenation of the two vector quantities of equal dimensions. The dimension of the MLP's output is the same as the output dimension of the message passing operations and has a linear activation. We fix the MLP to have two hidden layers with {\tt ReLU} activation and the same number of nodes as the output dimension. The network is implemented using the {\tt Pytorch-Geometric}~\cite{Fey/Lenssen/2019} package. Note that we have not performed any hyperparameter scan as part of this present, proof-of-concept study. We train the network for fifty epochs with a learning rate of 0.001 using the {\tt Adam}~\cite{kingma2014adam} optimiser. The training and validation losses are compared after each epoch to ensure that there is no overfitting or a premature termination of training. The epoch with minimum validation loss is used to infer the anomaly detection on the test dataset.

\section{Anomaly detection performance and results} 
\label{sec:results}
\begin{figure}[!t]
\includegraphics[width=0.48\textwidth]{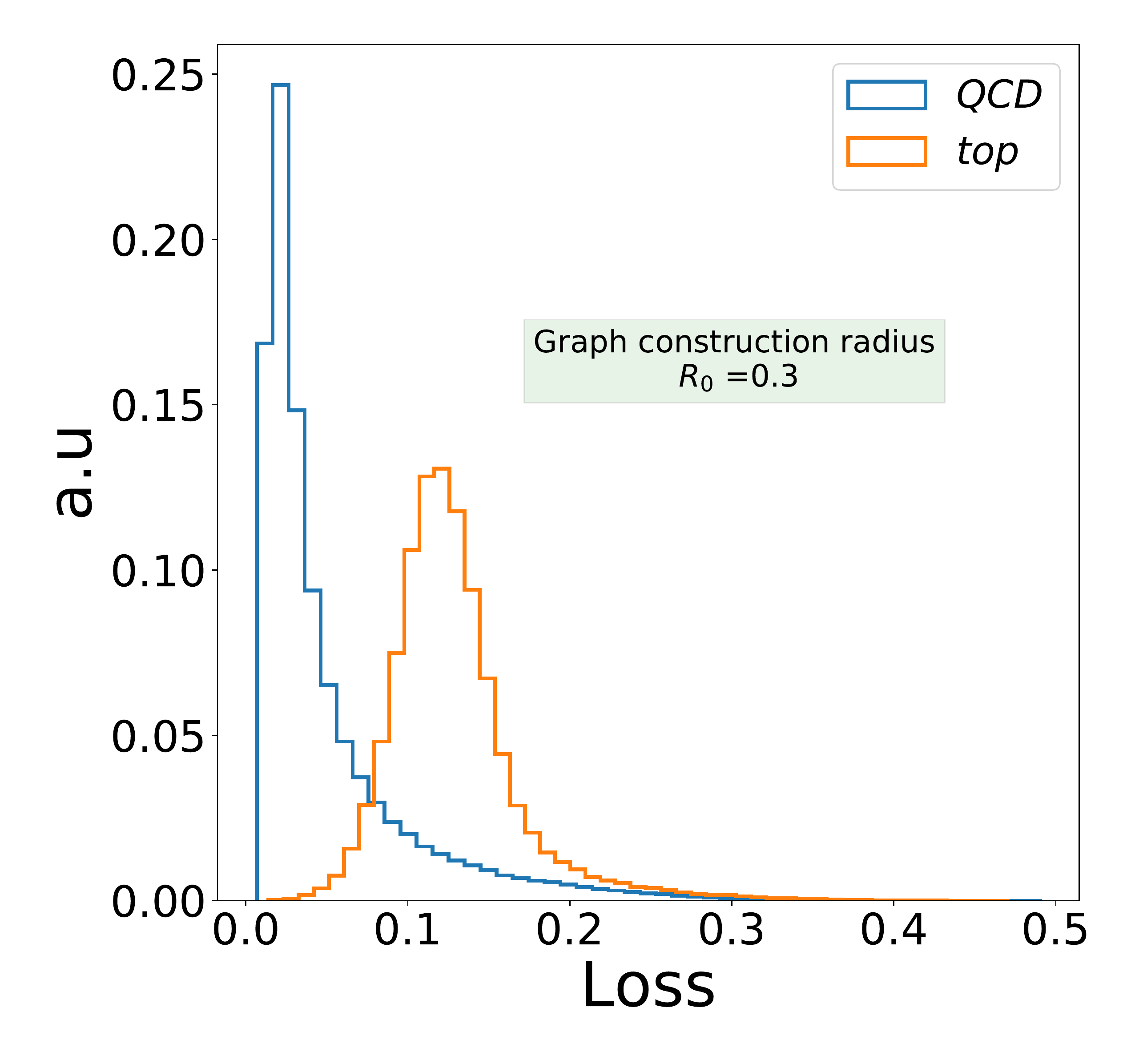}\hfill
\includegraphics[width=0.48\textwidth]{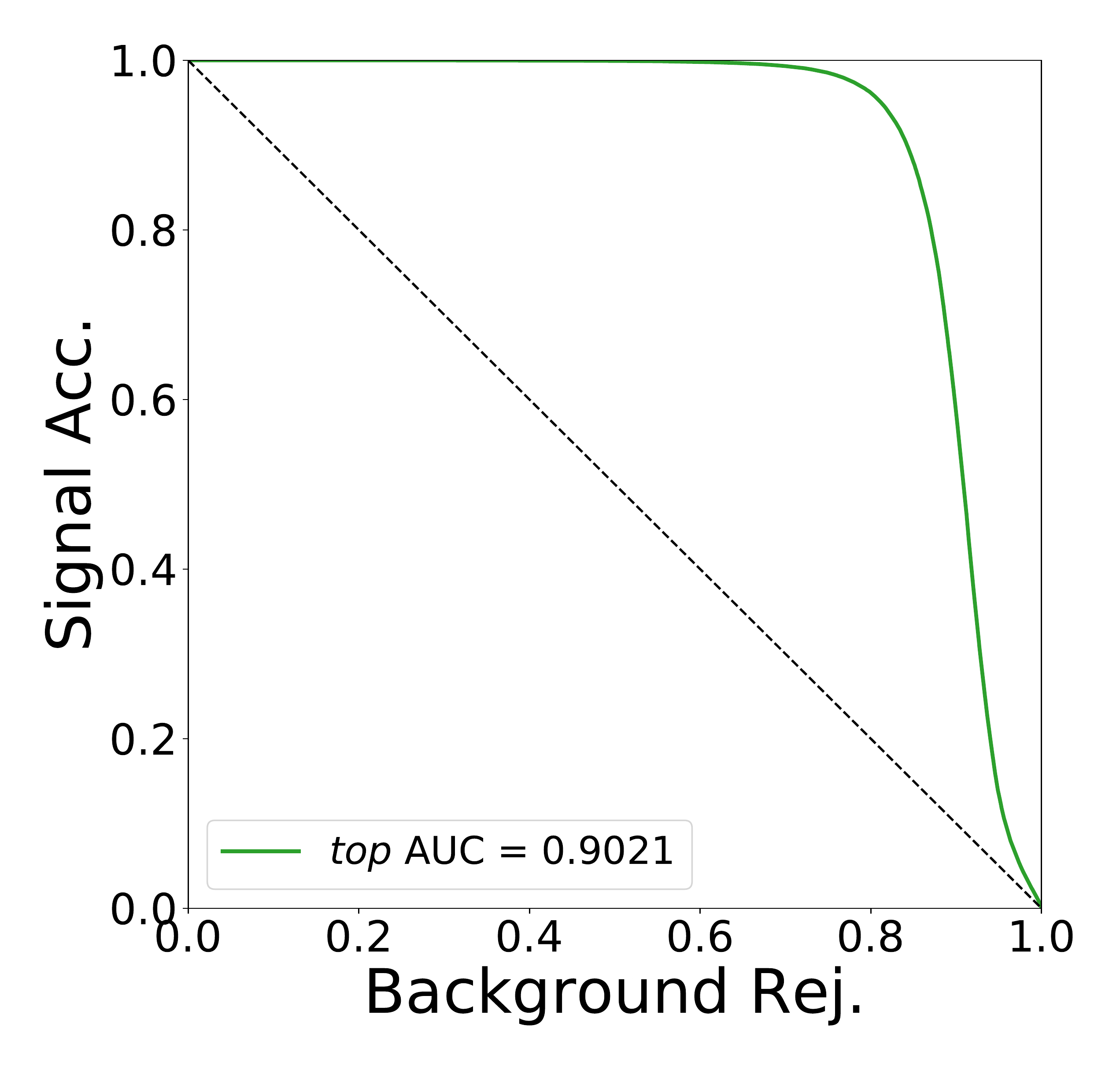}
\caption{The distribution of the loss function of an IRC safe graph autoencoder trained only with QCD jets with graph radius $R_0$=0.3\label{fig:loss}}
\end{figure}
We now discuss the performance of the designed IRC safe loss function in detecting anomalous jets when the network is trained only on the QCD background. We choose boosted top jets from the aforementioned public dataset as our benchmark. In Fig.~\ref{fig:loss} (left), we show the distribution of the loss function for the QCD and top jets (our inputs are the node features given in Eq.~\eqref{eq:node_feat}). As can be seen, the distributions of the loss function values for the QCD and top jets are significantly different, highlighting the capability of the architecture to detect anomalous jets in an IRC-safe way. The Receiver-Operator-Characteristic (ROC) curve and the Area Under the Curve (AUC) of 0.902 shown in Fig.~\ref{fig:loss} (right) confirm the good separation shown in the loss distribution, rivalling convolutional autoencoders~\cite{Farina:2018fyg,Heimel:2018mkt,Roy:2019jae,Finke:2021sdf} which also have AUCs close to such values (up to 0.93~\cite{Heimel:2018mkt} and 0.91~\cite{Finke:2021sdf}) on the same dataset. Although we did not perform a hyperparameter scan for this study, we observed a decrease in performance for a one-dimensional latent space. 

Top jets possess a different and hard kinematical structure that is typically not present in QCD jets. The ability to look into the soft and collinear splittings from the QCD shower evolution in an IRC safe way enables the network to access such information and the hard radiation pattern in a theoretically meaningful way. Modifications of the soft and collinear radiation patterns that are seeded by novel hard scales (see e.g. Refs.~\cite{Soper:2011cr,Englert:2011cg,Gerwick:2011tm,Gerwick:2012hq,Soper:2014rya,Prestel:2019neg} for a more traditional jet-based approach to this) are therefore consistently included in the anomaly detection performance. Therefore, when such non-QCD structures are present, the anomaly detection performance is considerably improved.

\begin{figure}[!t]
\centering
\includegraphics[width=.48\textwidth]{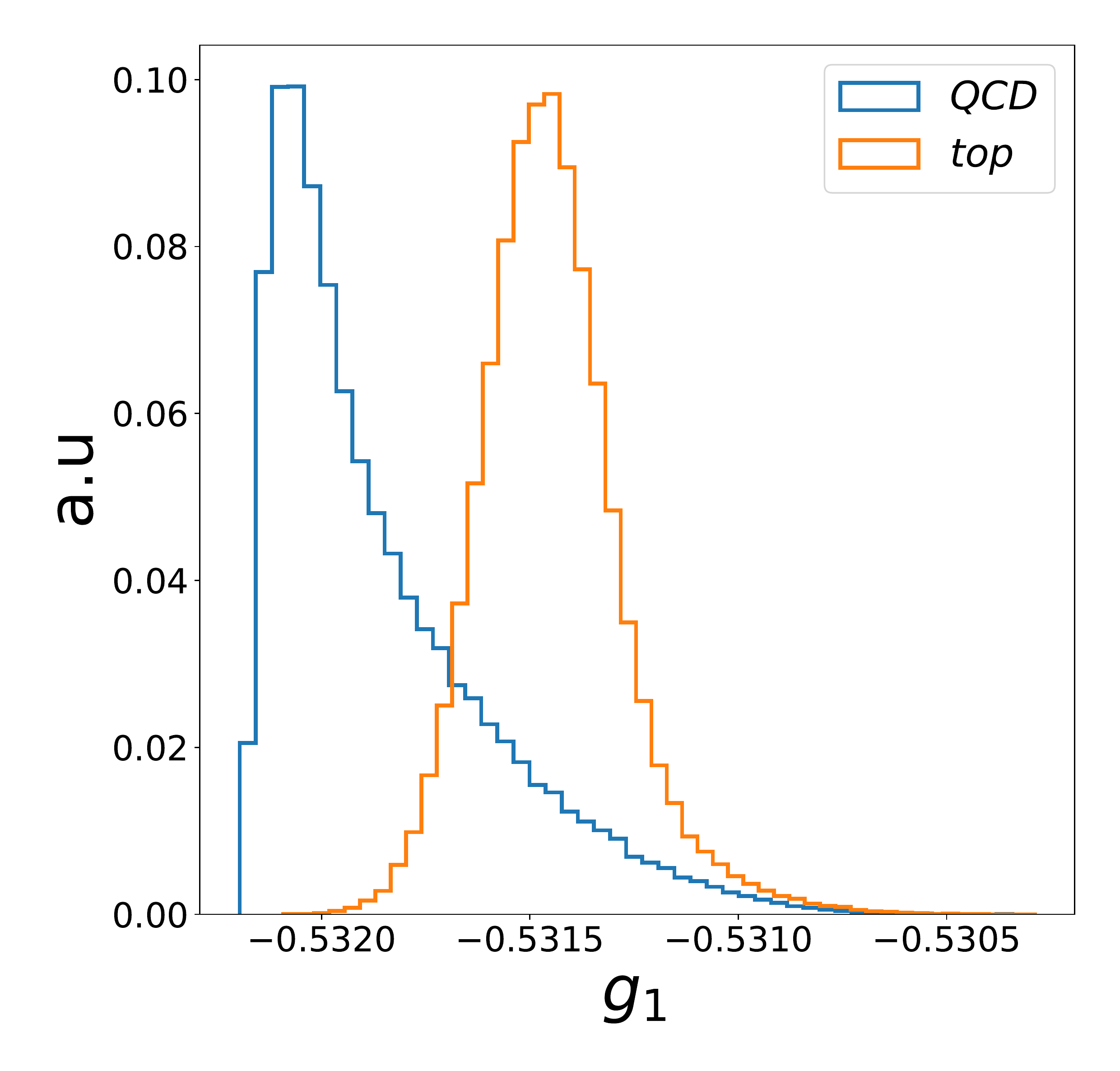}\hfill
\includegraphics[width=.48\textwidth]{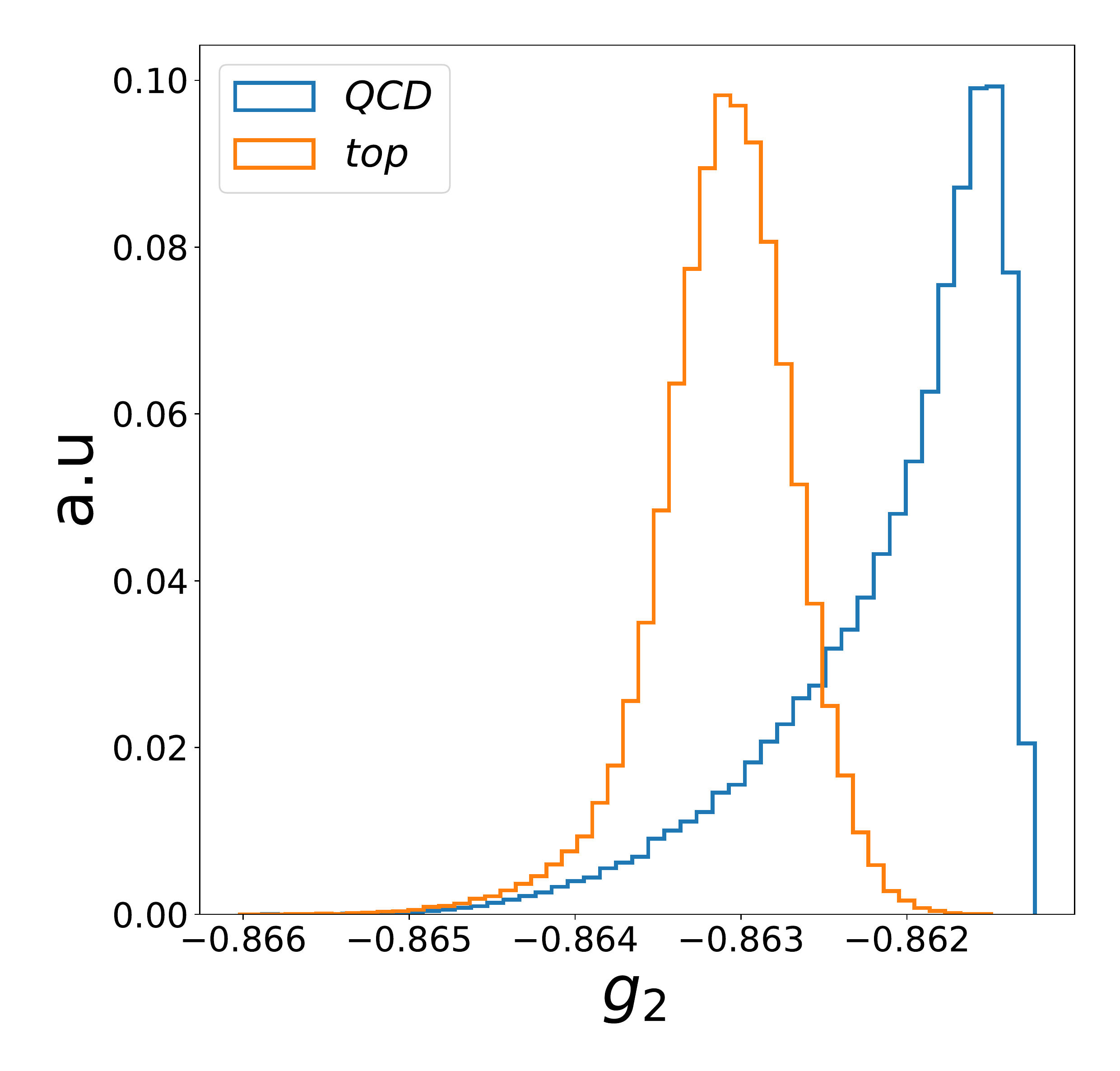}
\caption{The distribution of each dimension of the two-dimensional latent spaces obtained after an IRC safe graph readout given in Eq.~\eqref{eq:lat_graph}.\label{fig:lat_dist}}
\end{figure}
\begin{figure}[!t]
\centering
\includegraphics[width=.48\textwidth]{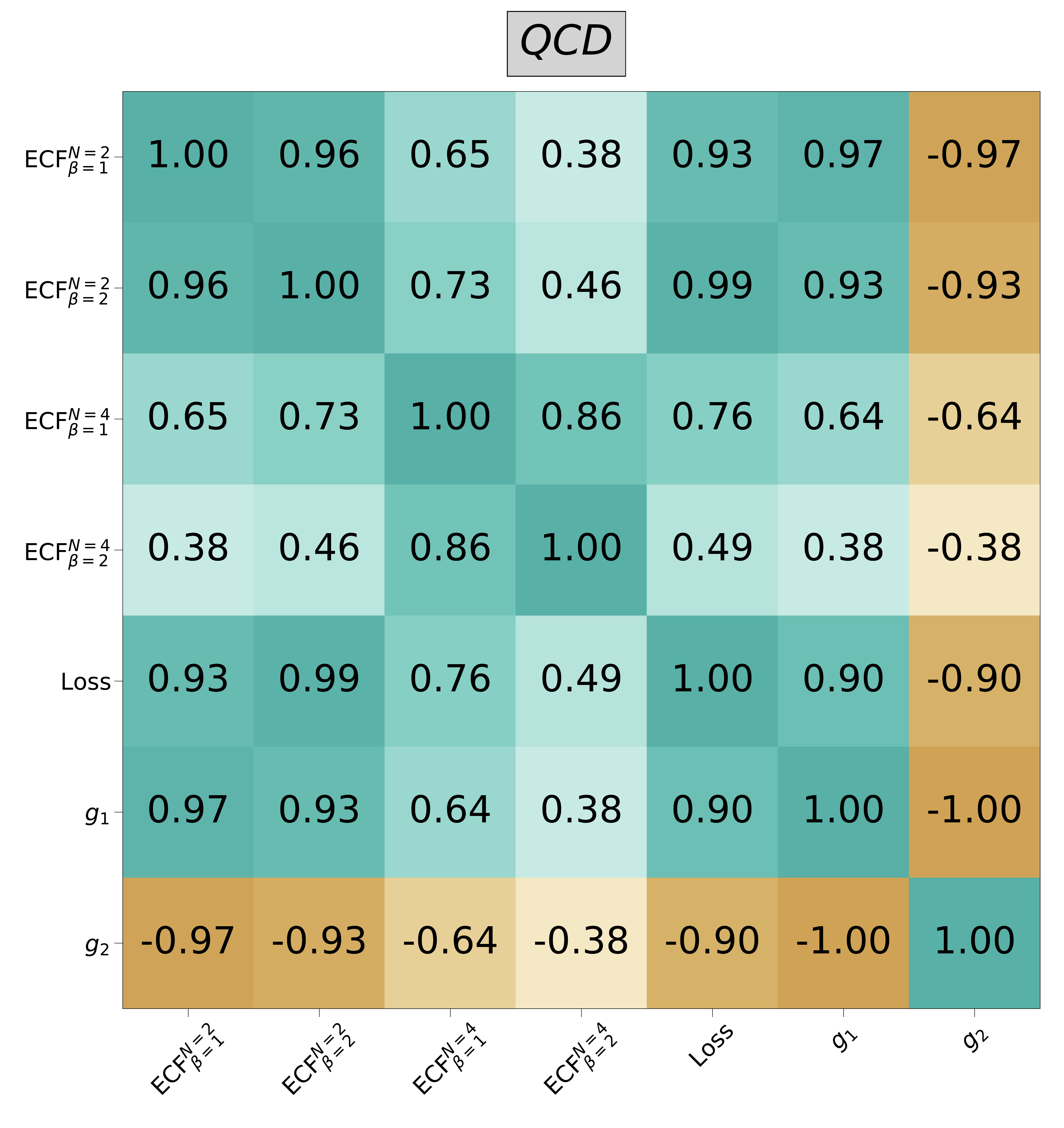}\hfill
\includegraphics[width=.48\textwidth]{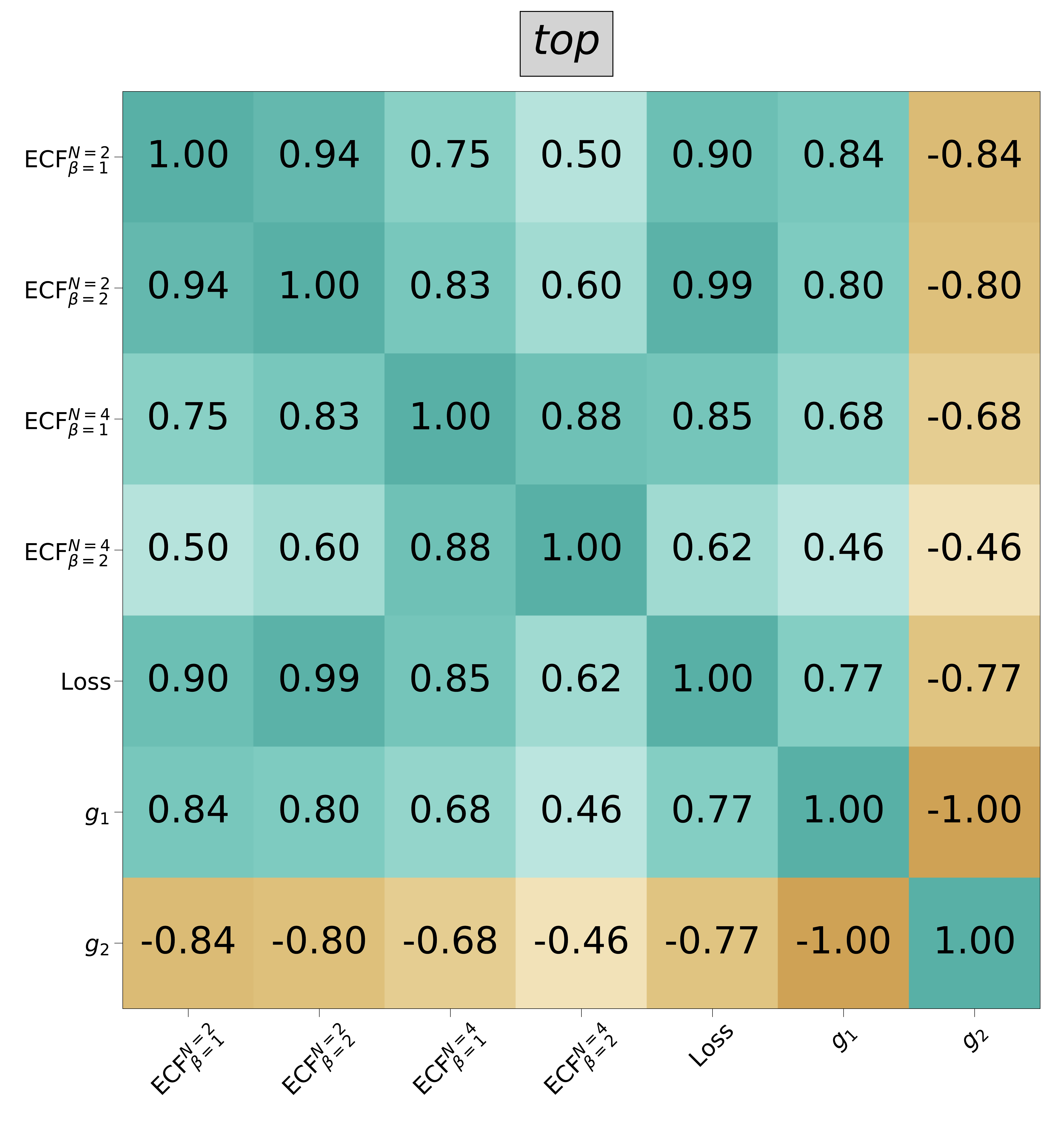}
\caption{The correlation of IRC safe loss (cf. Eq.~\eqref{eq:lNodeoss}) and latent dimension (obtained with Eq.~\eqref{eq:lat_graph}) is shown with the Energy Correlation Functions~\eqref{eq:ecf}. One can see a very high correlation of the ECFs with the variables obtained from the network, hinting at a close connection between them. \label{fig:corr}	}
\end{figure}

In light of these results, it is worthwhile to compare our autoencoder results to phenomenological aspects of QCD in jet substructure analyses. From the point of view of soft and collinear features, Energy Correlation Functions (ECF)~\cite{Larkoski:2013eya} are particularly relevant for such a comparison as we will motivate below. Furthermore, given that our autoencoder condenses the QCD information into the latent space in an IRC-safe way, it is interesting to see how it correlates with ECF observables.
To this end, we define
\begin{equation}
\label{eq:lat_graph}
\mathbf{g}=\sum_{a\in\mathcal{G}}\; z_a\;\mathbf{h}_a\,,	
\end{equation}  
where $\mathbf{h}_a$ are the latent node features. Similar to the graph readout in a classification scenario~\cite{Konar:2021zdg}, this is an IRC safe representation of the jet. The distribution of the individual components of the two-dimensional graph representation are shown in Fig.~\ref{fig:lat_dist}. The good performance of the autoencoder is reflected in the good separation in the latent space. The two latent space directions are, however, completely anti-correlated; see Fig.~\ref{fig:corr} (they are also highly correlated with the loss).
Thus, restricting ourselves to any of these three variables would be sufficient for the anomaly detection problem studied in this work. The loss would most likely be a better choice when one focuses on anomaly detection capabilities since it condenses the information of the two-dimensional latent space into a single quantity. On the other hand, any latent feature would be more suitable for applications demanding lower execution times, since in this case only the encoder needs to be evaluated during inference.



Moving on to the relation of the learned information with ECFs, we first define the ECFs as
\begin{equation}
\label{eq:ecf}
\text{ECF}(N,\beta)=\sum_{i_1<i_2<....i_{N-1}<i_N\in J}\;\left(\prod_{a=1}^{N} z_{i_a}\right)\;\left(\prod_{b=1}^{N-1} \prod_{c=b+1}^{N} \Delta R_{i_bi_c}^\beta\right)\,. 
\end{equation}
Focussing specifically on the case $N=2$, we obtain
\begin{equation}
\label{eq:ecf2}
\text{ECF}(2,\beta)=\sum_{j<i\in I_J}\;z_{i}\;z_{j}\;\Delta R_{ij}^\beta=\sum_{i=1}^{|J|}z_i\;\sum_{j=i+1}^{|J|}z_j\;\Delta R_{ij}^\beta\,, 
\end{equation} where $I_J$ is the index set of the constituent set $J$, and $z_i=p_T^i/(\sum_{k\in I_J} p_T^k)$. 
We can rewrite the expression as  
\begin{equation}
\text{ECF}(2,\beta)=\sum_{i=1}^{|J|}z_i\;H_i\quad,\quad H_i=\sum_{j=i+1}^{|J|}z_j\;\Delta R_{ij}^\beta\,.
\end{equation}
Therefore, the quantity $H_i$ can be regarded as a scalar node feature obtained from the message function  $\Delta R_{ij}^\beta =\Delta R_{ij}^\beta(\hat{p}_i,\hat{p}_j)$, with a weighted (by $z_j$) summed readout, while the sum over $i$ to get the ECF is similar to a graph readout operation on all the nodes (or constituents) of the jet.
Although the graph structure in the current case is the 2-combinatorial graph, such an analogy suggests that the features extracted by the EMPN are closely connected to ECFs.

This expectation is analysed in more detail in Fig.~\ref{fig:corr}, where we show the correlation of different order ECFs with each dimension $g_i$ of the latent graph readout and the loss function . There is a strong correlation between the 2-point ECFs and the network outputs, which decreases when considering the 4-point ECFs. This difference illustrates the close relation of the message passing architecture to the 2-point ECFs. 
The latent dimensions show a higher correlation for $\beta=1$ than $\beta=2$, while the opposite holds for the loss function. This may be due to the ReLU activation, which is essentially a linear function for all positive arguments, while the loss function's higher correlation to the quadratic ECFs may be due to the usage of the mean-squared error as $d(\hat{p}_i, \hat{\bar{p}}_i)$.

\section{Conclusions}
\label{sec:conc}
Infrared and collinear safety is not a luxury but an essential requirement to guarantee the theoretical consistency of particle physics collider data interpretations. The emerging and fast-developing area of anomaly detection should therefore incorporate IRC safety when analysing data at the LHC where QCD activity plays a dominant role. New heavy physics significantly deviates from QCD phenomenology, predominantly characterised by soft and collinear emissions. Reflecting the QCD expectation adequately helps isolate anomalies further; the ability to meaningfully interpolate into the soft and collinear regime is crucial for extending the reach of such techniques to lower scales. Despite this, IRC safety has not played an essential role in the implementation of anomaly detection. In this paper, we have placed IRC safety at the heart of anomaly detection for the first time by constructing a graph neural network autoencoder that employs Energy-Weighted Message-Passing, which gives rise to an IRC-safe architecture~\cite{Konar:2021zdg}. 

Graph neural networks are well-adapted approaches for isolating tell-tale correlations of final states~\cite{Dreyer:2020brq,Atkinson:2021jnj} and we find that our algorithm shows a high anomaly detection capability whilst having theoretically appealing properties. We have demonstrated this by injecting top jets as an anomaly and finding excellent discriminating sensitivity. While this partly results from the direct presence of a novel hard scale in the jet's substructure, additional sensitivity is accessed from a different soft and collinear shower pattern that accompanies the hard scale. To highlight this relation to well-studied observables in QCD phenomenology, we have shown a strong relation of the information encoded in our autoencoder's latent space with energy correlation functions. This motivates extending anomaly detection analyses using our framework to new physics scenarios of lighter BSM degrees of freedom, which we leave for future work.

\section*{Acknowledgements}
O.A. is supported by the UK Science and Technology Facilities Council (STFC) under grant ST/V506692/1.
A.B. and C.E. are supported by the STFC under grant ST/T000945/1. 
C.E. is also supported by the Leverhulme Trust under Research Project Grant RPG-2021-031 and the IPPP Associate Scheme.
P.K. and V.S.N. are supported by the Physical Research Laboratory (PRL), Department of Space, Government of India. M.S. is supported by the STFC under grant ST/P001246/1. Part of the computational work detailed in this paper was performed using the HPC resources (Vikram-100 HPC) and TDP project at PRL.


\bibliographystyle{JHEP}
\bibliography{references}

\end{document}